\documentclass[12pt]{aastex}

\shorttitle{mCV Accretion Flows \& Evolution}
\shortauthors{Norton, Butters, Parker \& Wynn}

\def\porb{P_{\rm orb}}
\def\pspin{P_{\rm spin}}

\begin{document}

\newcommand{\lta}{{\small\raisebox{-0.6ex}{$\,\stackrel
{\raisebox{-.2ex}{$\textstyle <$}}{\sim}\,$}}}
\newcommand{\gta}{{\small\raisebox{-0.6ex}{$\,\stackrel
{\raisebox{-.2ex}{$\textstyle >$}}{\sim}\,$}}}   

\title{The Accretion Flows and Evolution of Magnetic Cataclysmic Variables}

\author{A.J. Norton}
\affil{Department of Physics and Astronomy, The Open University, Walton Hall,
Milton Keynes MK7 6AA, U.K.}
\email{A.J.Norton@open.ac.uk}

\author{O.W. Butters}
\affil{Department of Physics and Astronomy, The Open University, Walton Hall,
Milton Keynes MK7 6AA, U.K.}
\email{O.W.Butters@open.ac.uk}

\author{T.L. Parker}
\affil{Department of Physics and Astronomy, University of Leicester, 
Leicester LE1 7RH, U.K}
\email{tp57@le.ac.uk}

\and

\author{G.A. Wynn}
\affil{Astronomy Group, University of Leicester, Leicester LE1 7RH, U.K.}
\email{gaw@astro.le.ac.uk}

\begin{abstract} 

We have used a model of magnetic accretion to investigate the accretion
flows of magnetic cataclysmic variables (mCVs). Numerical simulations 
demonstrate that four types of flow are possible: discs, streams, 
rings and propellers. The fundamental observable determining the accretion 
flow, for a given mass ratio, is the spin-to-orbital period ratio of the 
system. If IPs are accreting at their equilibrium spin 
rates, then for a mass ratio of 0.5, those with $\pspin/\porb \lta 0.1$ will 
be disc-like, those with $0.1 \lta \pspin/\porb \lta 0.6$ will be 
stream-like, and those with $\pspin/\porb \sim 0.6$ will be ring-like. The 
spin to orbital period ratio at which the systems transition between these 
flow types increases as the mass ratio of the stellar components decreases.

For the first time we present evolutionary tracks of mCVs which allow
investigation of how their accretion flow changes with time. As systems evolve 
to shorter orbital periods and smaller mass ratios, in order to maintain spin 
equilibrium, their spin-to-orbital period ratio will generally increase. As 
a result, the relative occurrence of ring-like flows will increase, and the 
occurrence of disc-like flows will decrease, at short orbital 
periods. The growing number of systems observed at high spin-to-orbital 
period ratios with orbital periods below 2~h, and the observational evidence 
for ring-like accretion in EX~Hya, are fully consistent with this picture.

\end{abstract}

\keywords{ accretion, accretion discs - binaries: close -
stars: magnetic fields.}

\section{Introduction}
\label{sec:intro}

Magnetic cataclysmic variable stars (mCVs) provide an excellent probe of 
the accretion process under extreme astrophysical conditions. They are 
interacting binary stars in which a magnetic white dwarf (WD) accretes material
from a late-type companion star via Roche lobe overflow. The WD has a large 
magnetic moment ($\mu_1 \sim 10^{32} - 10^{35}$~G~cm$^3$) which has a
wide ranging influence on the dynamics of the accretion flow. The mCVs fall 
into two distinct classes: the AM~Herculis stars (or polars) and the 
intermediate polars (IPs or DQ~Herculis stars); a comprehensive review 
may be found in Warner (1995).  The rotational periods of the WDs 
($\pspin$) in polars are generally locked to the orbital period 
($\porb$), whereas the WDs in IPs are asynchronous with $\pspin/\porb \approx 
0.01 - 0.6$. Polars contain the most strongly magnetic WDs and their 
synchronism is thought to come about due to the interaction between the 
magnetic fields of the two stars which is able to overcome the spin-up torque 
of the accreting matter (see e.g. King, Frank and Whitehurst 1991). IPs fill 
the parameter space between the strongly magnetic polars and the non-magnetic 
CVs. The accretion flows within IPs are known to take on a wide variety of 
forms, from magnetically confined accretion streams to extended accretion 
discs, similar to non-magnetic CVs. This variety has constantly perplexed 
efforts to understand these objects and is the subject of this work.

In an earlier paper (Norton, Wynn \& Somerscales 2004; hereafter paper 1)
we used a model of magnetic accretion to investigate the rotational
equilibria of mCVs. We showed that there is 
a range of parameter space in the $\pspin / \porb$ versus $\mu_1$ plane at 
which rotational equilibrium occurs. This allowed us to infer
approximate values for the magnetic moments of all known intermediate polars.
As a result we established that
the number of systems as a function of white dwarf magnetic moment is
distributed approximately according to $N(\mu_1) {\rm d}\mu_1 \propto
\mu_1^{-1} {\rm d}\mu_1$.
In that paper we noted that the spin equilibria correspond to a variety of
different types of accretion flow, including disc-like accretion at small
$\pspin / \porb$ values, stream-like accretion at intermediate $\pspin / \porb$
values, and accretion fed from a ring at the outer edge of the white dwarf
Roche lobe at higher $\pspin / \porb$ values. In this paper we investigate
these flows in a more systematic manner and examine how the accretion flow 
varies as a system evolves. In particular for the first time we determine the 
evolutionary tracks of IPs at a range of magnetic field strengths  and compare
these with the observed distribution of IPs.

\section{Magnetic Accretion Flows and the Magnetic Model}
\label{sec:magacc}

Full details of how the magnetic accretion flow may be characterized and 
how we model this may be found in paper 1. Briefly, we assume that material
moving within the binary system interacts with the local WD magnetic field
via a shear velocity-dependent acceleration. This is analogous to the 
assumption that the magnetic stresses are dominated by the magnetic tension
rather than the magnetic pressure, which will be valid in all but the 
innermost regions of the flow, close to the white dwarf surface. We thus 
write the magnetic acceleration as a coefficient ($k$) multiplied by the 
difference in velocity between the accreting material and the field lines 
($\mbox{\boldmath$v$}_{\perp}$). 
\begin{equation}
k(r) \mbox{\boldmath$v$}_\perp = \frac{1}{\rho(r) R_{\rm c}(r)}
\frac{B^2(r)}{4\pi} \mbox{\boldmath$\hat{n}$}
\end{equation}
where  the unit vector {\boldmath$\hat{n}$} is perpendicular to field lines, 
$\rho(r)$ is the local density of plasma and $R_{\rm c}(r)$ is the 
local radius of curvature of the field lines. 
$k$ therefore contains the details of the plasma-magnetic field interaction
and, as shown in paper 1, it scales as
\begin{equation}
k(r) = k_0 \left( \frac{r}{r_{\rm wd}} \right)^{-3}
\end{equation}
where $k_0$ is the parameter which is input to the modelling code and 
encodes both the magnetic field strength at the white dwarf surface
and the plasma density at the L1 point. A Gaussian distribution of $k_0$ 
is used in the model, with typically a standard deviation of 1\% around the 
mean value to represent the range of plasma densities in the flow. Larger 
values of this Gaussian width result in more `blurred' boundaries between 
regions displaying different accretion flows (see below) but do not alter the 
white dwarf equilibrium spin period for a given $k_0$.

The results we present here were obtained with a three-dimensional 
particle hydrodynamics code known as HyDisc, using an 
implementation of the model described fully in paper 1. The calculations are 
carried out in the full binary potential and include a simple treatment of 
the gas viscosity. Previous results obtained with HyDisc are described in
King \& Wynn (1999); Wynn, King \& Horne (1997) and Wynn \& King (1995).
The HyDisc code uses `packets' of plasma, injected at the L1 point.
The density and size scale of these packets change as they travel towards
the white dwarf and they interact individually with the 
magnetic field lines. This structure mimics the `blobby' accretion 
seen in (e.g.) AM Her (Heise et al. 1985; Hameury \& King 1988) which 
gives rise to large variations in the density of the flow at the white dwarf 
surface.

\section{Flow Results}

\subsection{Triple points in the $\pspin/\porb$ vs. magnetic moment plane}

In paper 1 we reported the accretion flows corresponding only to the 
equilibrium spin periods, as a function of orbital period and magnetic field
strength. Although mCVs are expected to remain close to equilibrium when
considered on long timescales, at a given instant a particular system
may be spinning up or spinning down. Indeed, Patterson (1994) predicted that
the period derivative of an IP in equilibrium will not be steady. The observed 
spin-up and spin-down rates in IPs typically correspond to much longer 
timescales ($\sim 10^9$~yr) than the timescale to reach spin equilibrium 
($\sim 10^7$~yr), indicating that systems are only exhibiting random excursions
away from equilibrium, probably driven by mass transfer fluctuations.

In order to explore the accretion flows corresponding to the full range of mCV 
parameter space, including behaviour away from spin equilibrium,  the magnetic 
model was run for each combination of parameters in a grid defined 
by the orbital period ($\porb = 80$min to 9h), spin period ($ \pspin = 
100$s to 5h), magnetic field strength ($k = (10^2 - 10^7) (2\pi / \porb) 
{\rm s}^{-1}$), and the mass ratio of the two stellar components ($q = 
M_2/M_1 = 0.2, 0.5, 0.9$). This corresponds to a range of white dwarf 
magnetic moments from about $10^{32}$~G~cm$^3$ to $10^{36}$~G~cm$^3$.
The secular mass transfer rate appropriate to the particular orbital period 
was assumed in each case. A model based on each combination of parameters was 
allowed to run to steady state before the nature of the resulting accretion was
examined. An atlas of these flows may be found in the PhD thesis by Parker 
(2005). Broadly speaking, each of the flows may be characterized as one of: 

\begin{itemize}
\item[{\bf propellers}] in which most of the material transferred
from the secondary star is magnetically propelled away from the system 
by the rapidly spinning magnetosphere of the WD.
\item[{\bf discs}] in which most of the material forms a circulating 
flattened structure around the WD, truncated at its inner edge by the WD 
magnetosphere where material attaches to the magnetic field lines before
accreting on the WD surface.
\item[{\bf streams}] in which most of the material latches onto the field
lines immediately and follows these on a direct path down to the WD.
\item[{\bf rings}] in which most of the material forms a narrow annulus 
circling the WD at the outer edge of its Roche lobe, with material 
stripped from its inner edge by the magnetic field lines before 
being channelled down to the WD surface.
\end{itemize}

We show in Figure 1 the results of analysing where the various flow types
occur, for systems with a mass ratio of $q=0.5$. Each panel is for a 
particular orbital period, and the $\pspin$
vs. $\mu_1$ plane is divided according to where each flow pattern is 
observed. We emphasize that the boundaries between the flow types are
generally rather blurred, and this blurring increases a little as the Gaussian 
spread of input $k_0$ values is increased. Nonetheless, the plane divides
into the regions shown, with the bold line marking roughly the locus of the 
equilibrium spin period in each case, as derived in paper 1. Clearly this
marks the boundary between accretion flow types that will generally spin-up
the WD (streams) and accretion flow types that will generally
spin-down the WD (propellers). Broadly speaking, if an asynchronous 
mCV finds itself in a region of parameter space where it is fed by a stream,
this will spin-up the white dwarf and so shift it downwards in the plane 
of Figure 1 towards the equilibrium line. Similarly, if an asynchronous 
mCV finds itself in a region of parameter space where the flow takes the 
form of a propeller, this will spin-down the white dwarf and so shift it 
upwards in the plane of Figure 1, again towards the equilibrium line. We note
that the accretion flow in the (possibly) magnetic system WZ Sge was modelled 
by Matthews et al. (2007) who derived a ring-like flow with a strong magnetic
propeller in that case. This system contains a rapidly spinning white dwarf
($P_{\rm spin} = 28$~s) and shows that other solutions are possible in
non-equilibrium situations such as this.

We note that disc and ring-like flows we see can each maintain the white dwarf 
close to spin equilibrium through a combination of accretion and ejection of 
material. As noted in paper 1, at equilibrium in the disc and ring-like 
flows, angular momentum from the WD is passed back to the accreting material, 
some of which is lost from the outer edge of the ring or disc to maintain 
equilibrium. Elsewhere in the parameter space at equilibrium, a 
stream/propeller combination is seen, which we have previously referred to 
as a weak propeller. As shown in Figure 2, close to the stream-disc-propeller 
triple point and the stream-ring-propeller triple point, the equilibrium flows 
are a combination of the various flow types. In each case at equilibrium the 
angular momentum accreted by the WD is balanced by an equal amount lost from 
the system via material which is magnetically propelled away from the WD. 
This, after all, is the definition of the equilibrium spin period. Hence, 
if real IPs sit at their equilibria they will exhibit accretion flows that are 
disc-like, stream-like or ring-like, each with a component of the flow that 
is propelled away.

As can be seen in Figure 1, both triple points move to smaller magnetic moments
as the orbital period decreases. However, for this mass ratio, the triple
points occur at the {\em same} spin-to-orbital period ratio at all orbital 
periods. In particular, the lower triple point is always close to 
$\pspin/\porb \sim 0.1$ whilst the upper triple point is always close to 
$\pspin/\porb \sim 0.6$. Hence, for a mass ratio of 0.5, if IPs are accreting 
at their equilibrium spin rates, those with $\pspin/\porb \lta 0.1$ will be 
disc-like, those with $0.1 \lta \pspin/\porb \lta 0.6$ will be stream-like, 
and those with $\pspin/\porb \sim 0.6$ will be ring-like. Equilibrium is not 
possible for $\pspin/\porb \gta 0.6$, until a system reaches synchronism 
(and is therefore a polar, exhibiting stream-fed accretion once more). 
Details of the synchronisation condition are given in paper 1.

\subsection{Changing the range of plasma density}

Changing the spread in the input $k_0$ value mimics changing the range of
plasma density throughout the flow. For the simulations described above, the 
$k_0$ values had a Gaussian distribution with a standard deviation of 1\% of
the mean value. Increasing this width to 10\% or 100\%
results in the changes to the flow shown in Figure 3. These simulations 
each correspond to a system with $P_{\rm orb} = 4$~hr and $q=0.5$ and
sit in the four regions of the $P_{\rm spin}/P_{\rm orb}$ vs. $\mu_1$ plane
identified above. In this case, the disc-like flow corresponds to a magnetic 
moment of $\sim 10^{33}$~G~cm$^3$ and a spin period of 1000~s, the stream-like
flow to $\sim 10^{34}$~G~cm$^3$ and 5000~s, the propeller-like
flow to $\sim 10^{35}$~G~cm$^3$ and 1000~s, and the the ring-like
flow to $\sim 2 \times 10^{36}$~G~cm$^3$ and 5000~s. As can be seen, the four
types of flow are still readily classified, and the effect of broadening the 
range of plasma densities in each model is minimal.

\subsection{Changing the mass ratio}

Changing the mass ratio of the stars in the system changes where the 
equilibrium spin period occurs. Figure 4 shows representative diagrams
for mass ratios of $q=0.2$, $q=0.5$ and $q=0.9$ for the case of a 4~h orbital
period. The same pattern of accretion flow behaviours is seen, but the 
triple points move to larger $P_{\rm spin}/P_{\rm orb}$ ratios and larger 
WD magnetic moments as the mass ratio decreases. 

King \& Wynn (1999) noted that mCVs have an equilibrium condition
specified by $R_{\rm co} \sim R_{\rm circ}$. Here $R_{\rm co}$
is the co-rotation radius, namely that at which matter
in local Keplerian rotation co-rotates with the magnetic field of the white 
dwarf, and $R_{\rm circ}$ is the circularization radius, namely that at which 
the specific angular momentum equals that of matter at the inner Lagrangian 
point. This in turn yields the condition
\begin{equation}
\frac{\pspin}{\porb} \sim (1+q)^2 (0.500 - 0.227 \log q)^6
\end{equation}
We identify this equilibrium with the lower triple point in Figures 1
and 4. For the three mass ratios examined here (i.e. $q = 0.2$, 0.5 and 0.9),
Equation 3 predicts spin-to-orbital period ratios of 0.118, 0.076 and 0.064
respectively. From our simulations,  the triple points are at spin-to-orbital
period ratios of 0.097, 0.083 and 0.063, in good agreement with the 
predictions.

King \& Wynn (1999) also noted another possible equilibrium, 
where $R_{\rm co} \sim b$, i.e. the distance from the white dwarf to
the inner Lagrangian point. This yields the condition
\begin{equation}
\frac{\pspin}{\porb} \sim (0.500 - 0.227 \log q)^{3/2}
\end{equation}
We identify this equilibrium with the upper triple point in Figures 1
and 4, as it indicates ring-like accretion flow confined to the outer edge
of the white dwarf's Roche lobe. For the three mass ratios examined here 
(i.e. $q = 0.2$, 0.5 and 0.9), Equation 4 predicts spin to orbital period 
ratios of 0.53, 0.43 and 0.36 respectively. From our simulations,  the triple 
points are at spin-to-orbital period ratios of 0.69, 0.56 and 0.49. The
slightly higher ratios observed probably reflect the fact that we observe 
ring-like structures to form just outside the white dwarf's Roche lobe, 
rather than at the edge of the lobe itself.

\section{Discussion}

\subsection{The observed distribution of intermediate polars}

Figure 5 shows the distribution of currently known mCVs in the $\pspin$ vs.
$\porb$ plane. The diagonal lines represent loci of constant spin-to-orbital
period ratio corresponding to the triple points at each of three mass ratio 
values. They therefore divide the plane into regions where different accretion
flows may be expected to occur. Regions below any of the three lines 
corresponding to the stream-disc-propeller triple points for mass ratios
of 0.2, 0.5 and 0.9 indicate where disc-like flows can occur; regions between
any of these three lines and the three lines 
corresponding to the stream-ring-propeller triple points for mass ratios
of 0.2, 0.5 and 0.9 indicate where stream-like flows can occur; and
the region around these upper three lines indicate where ring-like flows
will be most likely to occur. 

As can be seen, at least half of the IPs cluster around the 
region where the spin-to-orbit period ratio is in the range 
$0.05 \lta \pspin/\porb \lta 0.15$ which characterises the 
stream-disc-propeller triple point for plausible mass ratios. Assuming these
systems are accreting close to their equilibrium spin period, they are all
therefore likely to exhibit accretion flows which resemble the combination
disc-/stream-/propeller-like flow shown in the lower, centre panel of Figure 2.

There is a growing number of `EX Hya-like' systems below a 2~hr orbital period.
Many of these (e.g. SDSS~J023322.61+005059.5 (Southworth et al 2006), 
SDSS~J233325.92+152222.1 (Southworth et al 2007), DW Cnc (Patterson et al 
2004), V1025~Cen (Hellier, Wynn \& Buckley 2002) as well as EX~Hya itself) 
have high $\pspin/\porb$ ratios in the range 0.4 -- 0.7. As noted in paper 1  
they are likely to be characterised by ring-like 
accretion if they are at equilibrium. We note that this suggestion has recently
received considerable support from the spectroscopic observations of EX~Hya
presented by Mhlahlo et al (2007). The velocities they observed (in the 
range 500 -- 600 km s$^{-1}$) suggested 
that in this system material circulates the WD near to its Roche lobe,
and that accretion curtains are fed from a ring at this radius. This is 
exactly as predicted by our simulations. 

Finally, the systems at very small $\pspin/\porb$ ratios 
($\lta 0.01$) are likely to be either disc-like accretors
or, if they are out of equilibrium like AE Aqr, strong magnetic propellers.

Given that there are likely to be roughly equal numbers of magnetic CVs 
per decade of magnetic field strength (as derived in paper 1), we can 
comment on the expected distribution of accretion flows amongst the
observed population of IPs. Examining the distribution of accretion flow types 
in Figure 1, and assuming that all IPs are close to their spin
equilibria at all times, we can expect there to be relatively more disc-like
accretors at long orbital periods, and relatively more ring-like accretors
at short orbital periods. The number of stream-like accretors is likely to
be roughly constant at all orbital periods, as the region between the two
triple points occupies about one and a half decades of magnetic moment in 
each panel of Figure 1.

\subsection{The evolution of intermediate polars}

The observed distribution of IPs in the spin period / orbital period plane
is a result of observing systems with a range of magnetic field strengths 
at different stages in their evolution. Assuming they remain close to 
their equilibrium spin periods at all times, we can investigate how the 
observed distribution may be understood in terms of our results.

As IPs evolve, like all CVs, their mass ratio ($q=M_2/M_1$) will decrease
and they will move to shorter orbital periods
as they lose angular momentum via a combination of magnetic braking and
gravitational radiation. As a system evolves in this way, if the magnetic 
locking torque ($\propto \mu_1 \mu_2 /a^3$) exceeds the accretion torque, it 
may synchronize and emerge as a polar. As noted in paper 1, the reason we see 
several intermediate polars at short orbital periods may be because their
secondary stars have weak magnetic field strengths and so the magnetic locking
torque is ineffective. Alternatively, the magnetic fields of the two stars
may be mis-aligned in some way so as to minimize the effectiveness of this
mechanism.

In order to investigate the variation in spin period and accretion flow as
intermediate polars evolve, we took a typical theoretical evolutionary track 
of a cataclysmic variable, and followed this as it evolves to the orbital 
period minimum. In this evolutionary track, the WD had a constant mass of 
$1~{\rm M}_{\odot}$, and the mass ratio decreased from $q=1.18$ at $\porb=9$~h 
to $q=0.11$ at $\porb=80$~m. The mass accretion rate at each instant was
determined from the evolutionary model. We ran accretion flow simulations for
three different WD magnetic moments, namely $\mu_1 = 10^{32}$~G~cm$^3$ 
(i.e. $B_{\rm wd} = 0.6$~MG), $\mu_1 = 10^{33}$~G~cm$^3$ (i.e. $B_{\rm wd} = 
6$~MG), and $\mu_1 = 10^{34}$~G~cm$^3$ (i.e. $B_{\rm wd} = 60$~MG). For
each field value, we determined the equilibrium spin period of the white 
dwarf, following the same method as used in paper 1, at a range of orbital 
periods along the evolutionary track.

The results of this are shown in Figure 6, assuming that the systems under
study do not synchronize. In order to maintain spin equilibrium as a system 
evolves to shorter orbital periods, the WD spin period will generally change 
in the manner shown. Note that since both the mass ratio and orbital period
of the system vary continuously as the system evolves, the progress of the 
system cannot be easily tracked across the panels of Figure 1 or 4. In 
particular, since a decrease in orbital period causes the triple points to 
move to smaller magnetic moments, whilst a decrease in mass ratio causes the
triple points to move to larger magnetic moments, the behaviour of a given
system is not easy to predict. Nevertheless, it is apparent from our 
simulations that as systems evolve, their spin-to-orbital period ratios will 
generally increase and their accretion flows will become less 
disc-like and more stream-like. By the time they have crossed the period gap, 
the accretion flows are likely to be ring-like, and systems (if not 
synchronized) will appear similar to EX Hya with a large spin-to-orbital 
period ratio. As noted earlier, this has been supported by the recent 
observations of EX~Hya presented by Mhlahlo et al (2007) which show evidence
for a ring-like accretion flow. Furthermore, the 
growing number of systems discovered with high spin-to-orbital period
ratios at short orbital periods, as noted earlier, provides additional support
for the picture we have outlined.

We also note that Cumming (2002) has suggested that the relatively high 
accretion rates in IPs may overcome ohmic diffusion, such that magnetic flux 
is advected into the interior of the white dwarf, reducing the surface magnetic
field strength. This effective burying of the white dwarf magnetic field would 
make IPs appear less magnetic than they really are. Under this scenario, when
IPs emerge below the period gap after magnetic braking has presumably turned
off, their accretion rates will be substantially lower, and their `true'
magnetic field strengths might be expected to emerge. With a higher effective
magnetic moment, at a shorter orbital period, systems will be in spin 
equilibrium further to the right along the tracks in Figure 1 or 4, so making
a ring-like accretion flow even more likely for those systems that have
not synchronized to become polars. In terms of Figure 6, one can imagine a
system jumping from a lower magnetic field track to one corresponding to a 
higher magnetic field strength, as it emerges below an orbital period of 2~h.
However, we also note that Cumming's suggestion, which was extrapolated
from the case of accreting neutron stars, must be treated with caution as the 
timescale of the Rayleigh Taylor instability in the upper layers of an 
accreting WD differs substantially from that in neutron stars (Romani 1990). 
This may mean that the flows in WDs are not so easily buried after all. 
Furthermore, nova outbursts in accreting WDs will clear away much of the 
accreted layer, so helping to restore the field of the WD.

\section{Conclusions} 

Using a three-dimensional particle hydrodynamical model of magnetic 
accretion, we have demonstrated that broadly four types of accretion flow are 
possible in mCVs: discs, streams, rings and propellers. We have shown that 
the equilibrium spin periods in asynchronous mCVs, for a given orbital period 
and magnetic moment, occur where the flow changes from a type characterised 
by spin-up (i.e. stream-like) to one characterised by spin-down (i.e. 
propeller-like). As a result, the plane of WD spin period versus WD magnetic 
moment divides into four regions, one for each type of accretion flow, and 
contains a pair of triple points at which stream-disc-propeller and 
stream-ring-propeller flows co-exist. The first of these corresponds to when 
the co-rotation radius is equal to the circularization radius, and the second 
is when the co-rotation radius is equal to the distance from white dwarf to 
the L1 point. Changing the orbital period does not alter the spin-to-orbital 
period ratios at which these triple points occur, although they do move to 
smaller WD magnetic moments as the orbital period decreases. If IPs are 
accreting at their equilibrium spin rates, then for a mass ratio of 0.5, 
those with $\pspin/\porb \lta 0.1$ will be disc-like, those with $0.1 \lta 
\pspin/\porb \lta 0.6$ will be stream-like, and those with $\pspin/\porb \sim 
0.6$ will be ring-like.  In each case, at equilibrium some material is also 
propelled away from the system to maintain angular momentum balance. 
Decreasing the mass ratio increases the $\pspin/\porb$ ratio at the 
stream-disc-propeller and stream-ring-propeller triple points and also 
increases the WD magnetic moments at which they occur.

At long orbital periods, disc-like accretion flows are likely to be 
predominant, whilst at short orbital periods, the number of systems displaying 
ring-like accretion flows will increase. The relative number of systems 
displaying stream-like accretion flows is predicted to be roughly constant at 
all orbital periods. As IPs evolve to shorter orbital periods and smaller 
mass ratios, in order to maintain spin equilibrium, their spin-to-orbital
period ratios will generally increase. Those systems at short orbital periods 
that avoid syncronization are likely to appear similar to EX Hya, with a large 
spin-to-orbital period ratio and a ring-like accretion flow, as recently 
seen by Mhlahlo et al (2007).

\clearpage

\begin{figure}
\includegraphics[angle=-90,scale=0.6]{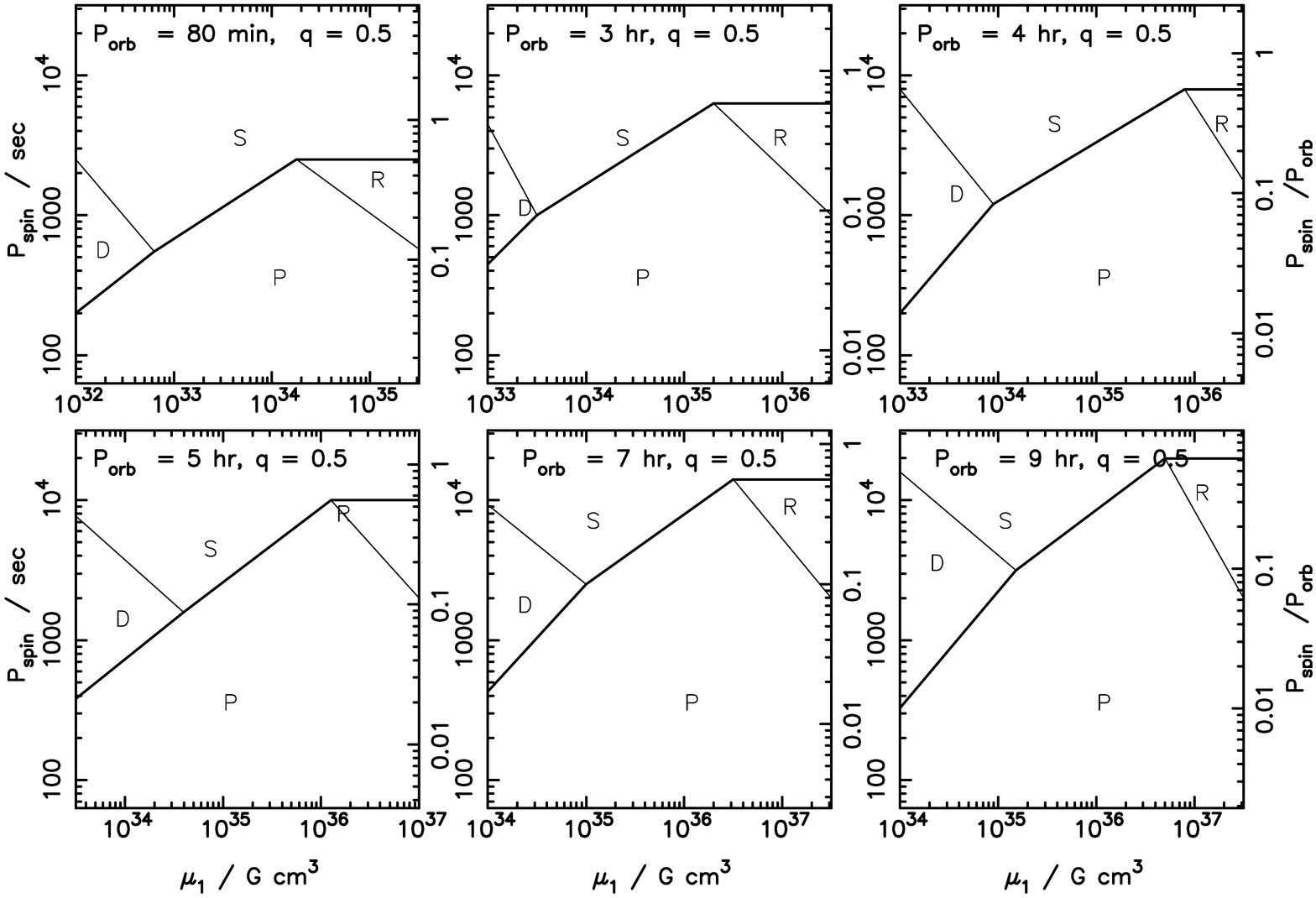}
\caption{The distribution of accretion flow types as a function of 
orbital period, in the spin period vs. magnetic moment plane, at a mass
ratio of $q=0.5$. The right 
hand axes show the spin to orbital period ratio in each case. Approximate 
regions within which each type of flow is seen are delineated as shown, where 
`D' stands for disc accretion, `S' for stream accretion, `R' for ring 
accretion and `P' for propeller flow. The thick line shows the approximate 
locus of the 
equilibrium spin period in each case and marks the boundary between
accretion flows that spin-up the WD and those which cause it to spin-down.
Note that the horizontal scale shifts between the panels to enable
us to plot the parameter space investigated at each orbital period. }
\end{figure}

\clearpage


\begin{figure}
\includegraphics[angle=-90,scale=0.7]{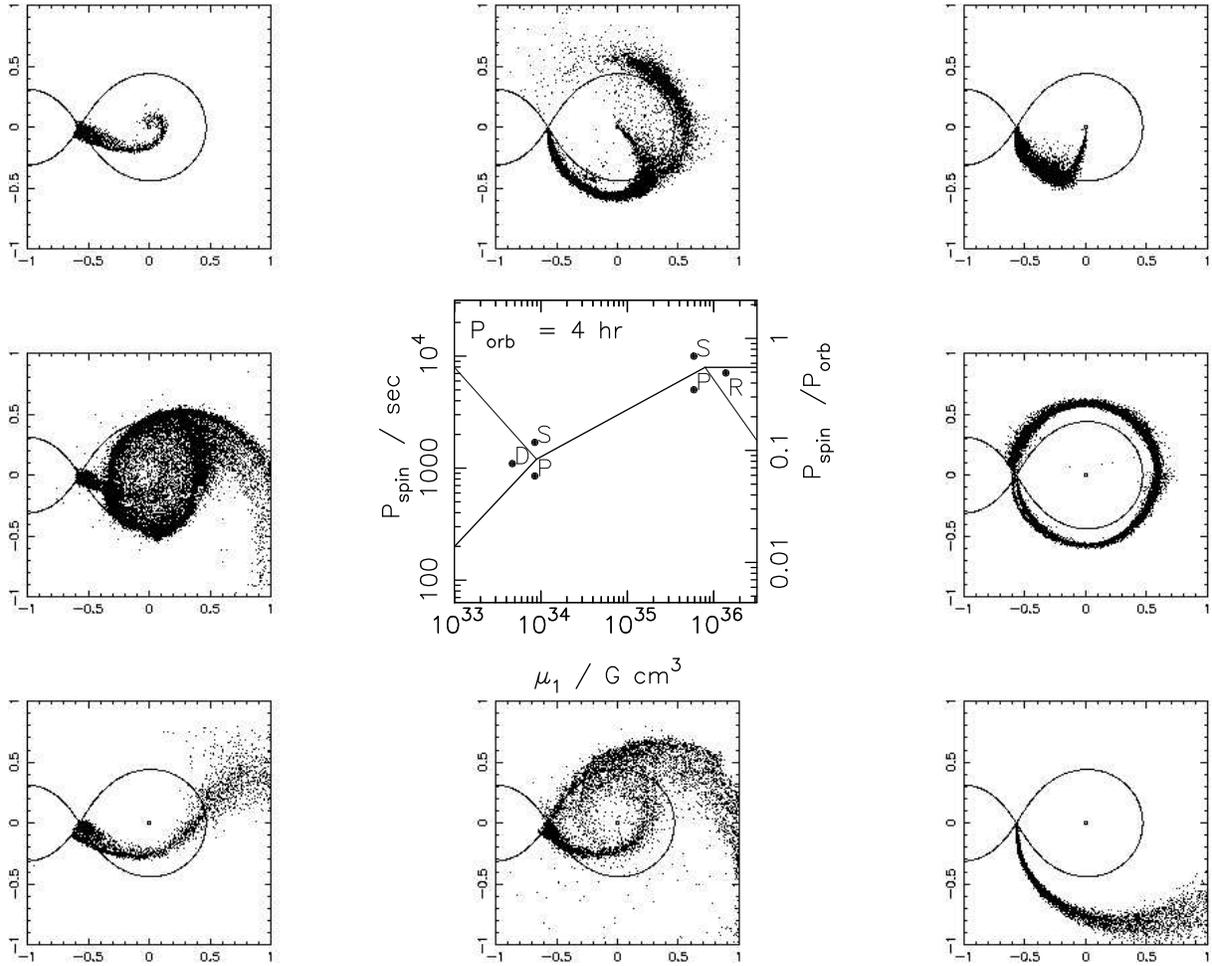}
\caption{The variation of the accretion flow in the vicinity
of the boundaries between the different flow types. The models shown are all
for $\porb$ = 4h and $q=0.5$. The panels on the left show flows in the 
vicinity of the 
stream-disc-propeller triple point, indicated by `S', `D' and `P' on the 
central panel, whilst the panels on the right show flows in the vicinity of 
the stream-ring-propeller triple point, indicated by `S', `R' and `P' on the 
central panel. The panel at the bottom, centre is the accretion flow {\em at}
the stream-disc-propeller triple point and shows characteristics of all three 
flows at an equilibrium spin period of $\sim 1200$~s for $\mu_1 \sim  
10^{34}$~G~cm$^3$. The panel at the top, centre is the accretion flow 
{\em at} the stream-ring-propeller triple point and shows characteristics of 
all three flows at an equilibrium spin period of $\sim 8000$~s for 
$\mu_1 \sim 5 \times 10^{35}$~G~cm$^3$.  }
\end{figure}

\clearpage

\begin{figure}
\includegraphics[angle=-90,scale=0.6]{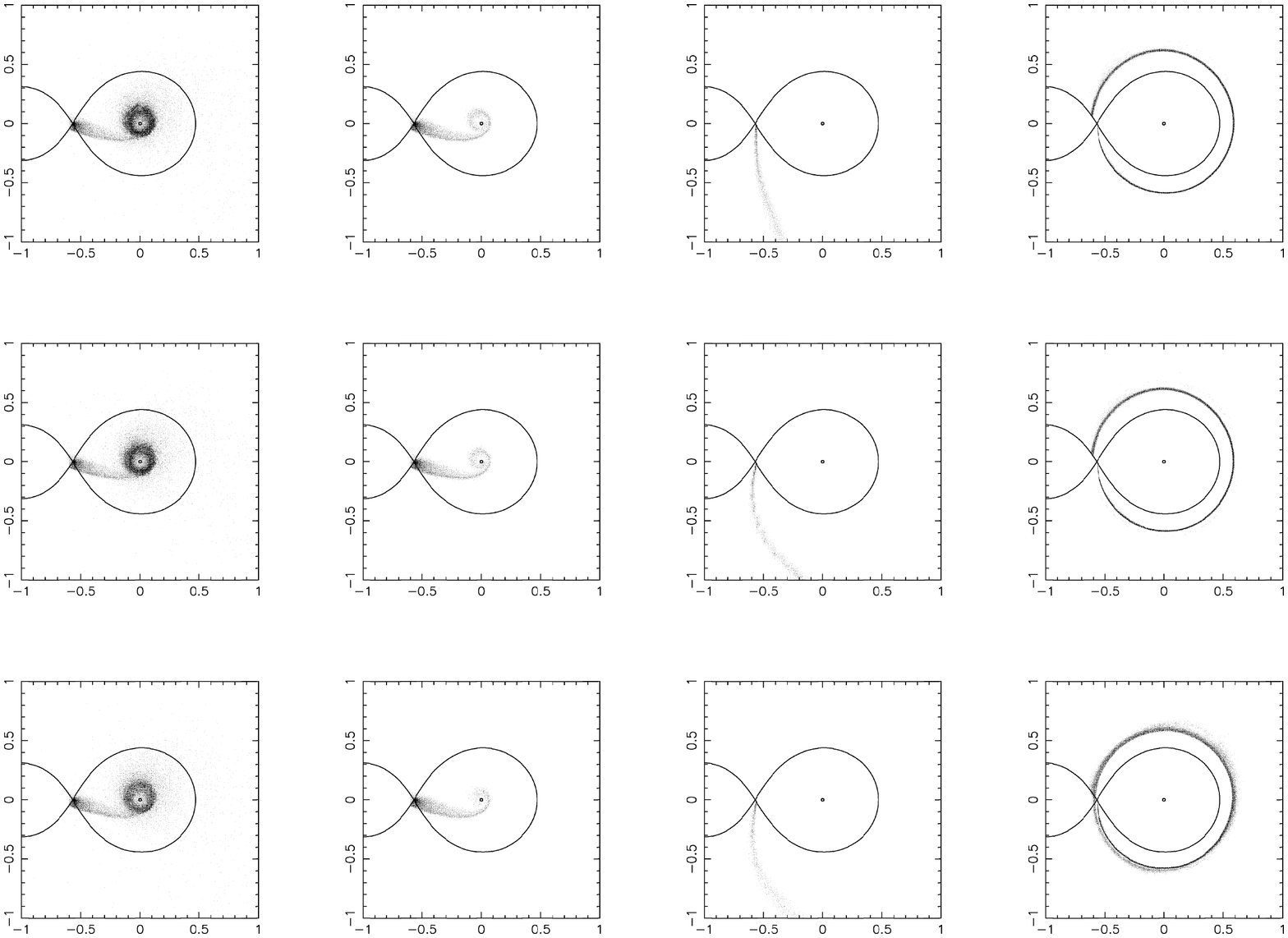}
\caption{Accretion flows for different ranges of plasma density. All the flows
shown here are for an orbital period of $P_{\rm orb} = 4$~h and a mass ratio
of $q=0.5$. In the top 
row, the standard deviation of the Gaussian distribution of $k_0$ values is 
1\% of the mean value, in the second row it is 10\% and in the third row it 
is 100\%. The disc-like flows (first column) are for a magnetic moment of
$\mu_1 \sim 10^{33}$~G~cm$^3$ and a spin period of $P_{\rm spin} = 1000$~s, the
stream-like flows (second column) are for $\sim 10^{34}$~G~cm$^3$ and 
5000~s, the propeller-like flows (third column) are for $\sim 10^{35}$~G~cm$^3$
and 1000~s, and the ring-like flows (fourth column) are for $\sim 2 \times 
10^{36}$~G~cm$^3$ and 5000~s. }
\end{figure}

\clearpage

\begin{figure}
\includegraphics[angle=-90,scale=0.6]{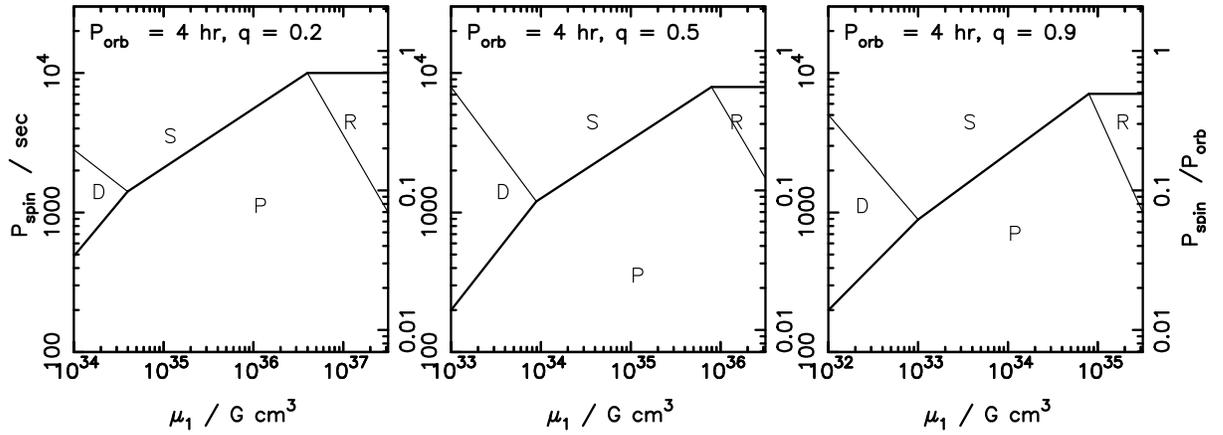}
\caption{The distribution of accretion flow types for mass ratios of 
$q=0.2$, $q=0.5$ and $q=0.9$ at an orbital period of $\porb = 4$~h, in the 
spin period vs. magnetic moment plane. The right 
hand axes show the spin to orbital period ratio in each case. `D' stands
for disc accretion, `S' for stream accretion, `R' for ring accretion and
`P' for propeller flow. The thick line shows the approximate locus of the 
equilibrium spin period in each case and marks the boundary between
accretion flows that spin-up the WD and those which cause it to spin-down. }
\end{figure}

\clearpage

\begin{figure}

\includegraphics[angle=-90,scale=0.6]{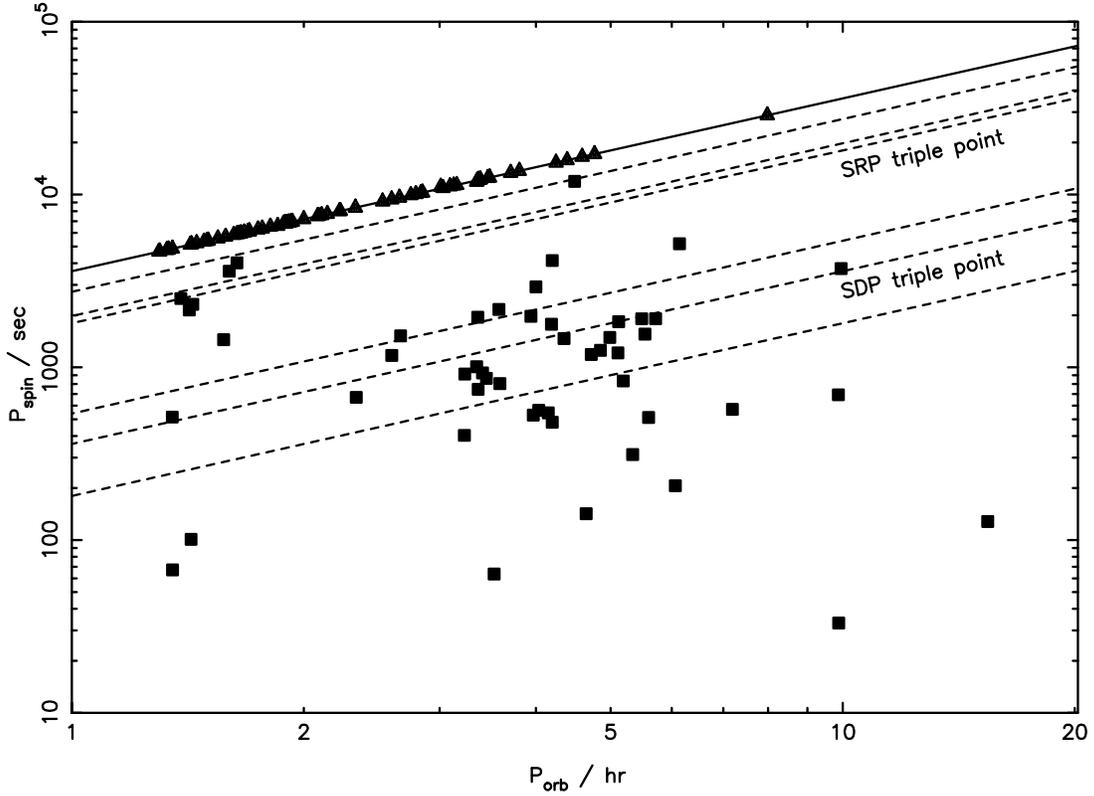}
\caption{The known mCVs distributed throughout $\pspin$ vs $\porb$ 
parameter space. Polars are indicated by triangles, and IPs by squares. The
lower set of three diagonal lines corresponds to the spin-to-orbital period 
ratio of the stream-disc-propeller triple point at mass ratios of 0.2, 0.5 
and 0.9 (top to bottom). The upper set of three diagonal lines correspond to 
the stream-ring-propeller triple point for the same mass ratios.}
\end{figure}

\clearpage

\begin{figure}

\includegraphics[angle=-90,scale=0.6]{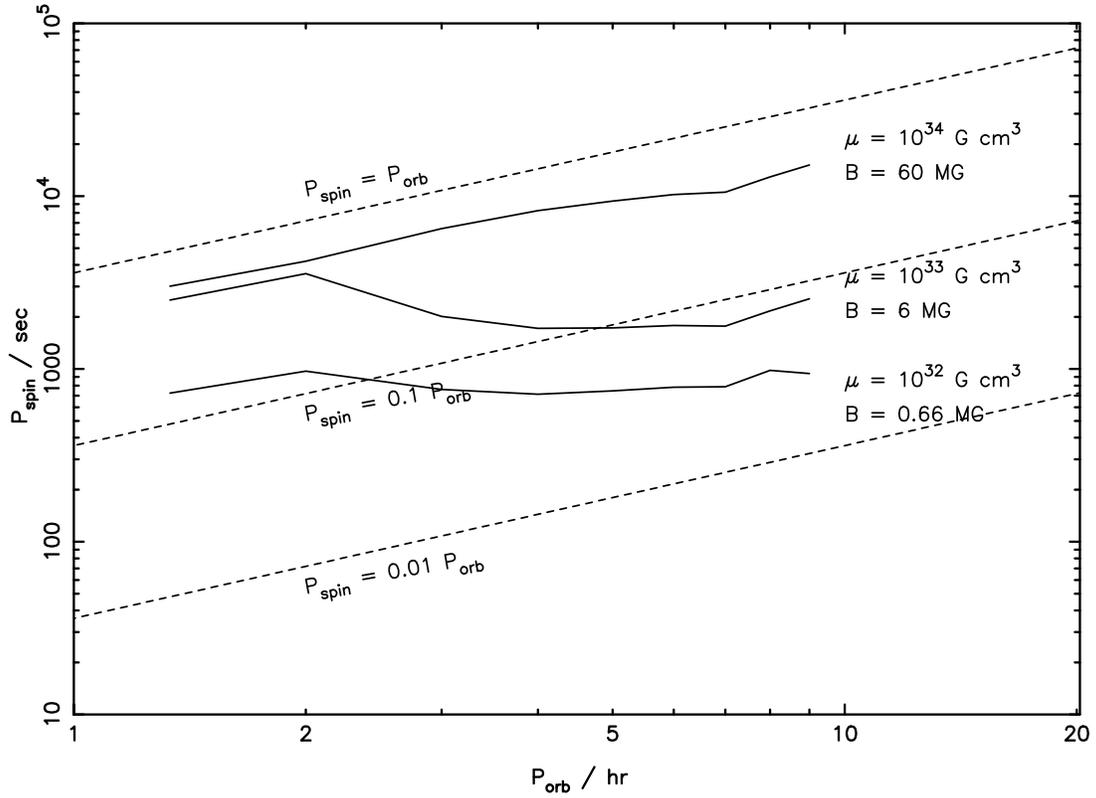}
\caption{The evolution of a magnetic CV with a solar mass white dwarf whose
magnetic moment is $10^{34}$~G~cm$^3$, $10^{33}$~G~cm$^3$, or 
$10^{32}$~G~cm$^3$, assuming that it does not synchronize. Evolution will
proceed from right to left in each case. Diagonal dashed
lines indicate tracks of constant spin-to-orbital period ratio.}
\end{figure}

\end{document}